# Three-Dimensional Ionisation, Dust RT and Chemical Modelling of Planetary Nebulae

B. Ercolano[1], M.J. Barlow[1], P.J. Storey[1]
[1]Dept. of Physics and Astronomy, University College London
, Gower St, London WC1E 6BT, UK

May 14, 2006

**Abstract**

The assumption of spherical symmetry is not justified for the vast majority of PNe. The interpretation of spatially-resolved observations cannot rely solely on the application of 1D codes, which may yield incorrect abundances determinations resulting in misleading conclusions. The 3D photoionisation code MOCASSIN (Monte CAarlo SimulationS of ionised Nebulae) is designed to remedy these shortcomings. The 3D transfer of both primary and secondary radiation is treated self-consistently without the need of approximations. The code was benchmarked and has been applied to the study of several PNe. The current version includes a fully self-consistent radiative transfer treatment for dust grains mixed within the gas, taking into account the microphysics of dust-gas interactions within the geometry-independent Monte Carlo transfer. The new code provides an excellent tool for the self-consistent analysis of dusty ionised regions showing asymmetries and/or density and chemical inhomogeneities. Work is currently in progress to incorporate the processes that dominate the thermal balance of photo-dissociation regions (PDRs), as well as the formation and destruction processes for all the main molecular species.

## 1 3D photoionisation models of spatially resolved PNe

The accurate quantitative interpretation of spectroscopic observations of a nebula requires the solution of the radiative transport (RT) problem, via the application of a photoionisation code. However such simulations are only preferable over empirical techniques if the models are sufficiently well constrained, given the degeneracy of the ionisation parameter. The geometry and density distribution of the object studied plays a major role in the resulting ionisation and temperature structure. A good constraint of the density distribution may be provided by narrow-band images, which are readily available for many objects in e.g. H$\alpha$, [O III], [N II], etc. A realistic 3D model should be able to match all available constraints, including projected 2D emission line maps. Clearly, there is still some uncertainty due to the loss of one dimension, but this can be resolved if velocity information is also available. We have shown, in the case of NGC 7009, how the oversimplifying assumption of spherical symmetry may lead to the wrong abundances being determined from spatially resolved observations of extended objects (see Goncalves et al. 2006 and poster at this conference). As an example, we present some recent calculations performed with the MOCASSIN code (Ercolano et al. 2003) of the ionisation structure of the young PN MyCn 18 (Ercolano, Neal and Sahai, in preparation). For this the 3D emissivity grids were rotated to the (independently derived) appropriate viewing angle and projected onto a plane normal to our line of sight (see Figure 1 of Ercolano, these proceedings). The images are compared here to narrow-band Hubble images (Sahai et al., 1999).

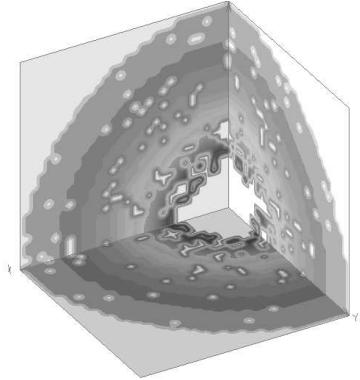

Figure 1: Cross-section slices showing a sample clumpy density distribution

## 2 Dust radiative transfer

The presence of dust grains in ionised environments affects the radiative transfer, as the grains compete with gas for the absorption of UV photons, as well as being heated by nebular resonance line photons. These processes can only be treated properly by incorporating the scattering, absorption and emission of radiation by dust particles, which are mixed with the gas. Moreover, the accurate determination of dust temperatures and SEDs can only be achieved by treating discrete grain sizes and different species separately. The dust and gas energetics are further coupled by a host of microphysical process, which include photo-electric emission from dust grains heating the gas and dust-gas collisions, which is a heating process for the dust and cooling process for the gas. All of the above has been fully achieved in the new version of MOCASSIN, which can also be used as a dust-only RT code and has been rigorously tested against established dust-only codes (Ercolano et al. 2005). While a number of Monte Carlo dust radiative transfer codes already exist, MOCASSIN is currently the only code availiable to treat the 3D dust radiative transfer within a photoionised nebula, providing the perfect tool for the modelling of non-spherical and/or inhomogeneous dusty photoionised environments.

# 3 Self-consistent PDR models

Extended PDRs have been detected around many young PNe, both through imaging and spectroscopy (e.g. NGC 7027). Additionally, density enhancements, such as the cometary globules observed in the Helix Nebula, may be opaque to ionising radiation, creating small-scale PDRs in shadows (e.g. Speck et al., 2002). MOCASSIN can model knots and small scale structures through the use of multi-grids of different spatial resolution. This new feature has already been applied to the modelling of inhomogeneous supernovae shells, as shown in Figure 1 (Ercolano et al., in preparation). We are currently working to incorporate the processes that dominate the heating and cooling within PDRs, as well as the formation and destruction processes for all the main species. We have already included all the relevant opacity contributions in the PDR region and are currently in the process of developing algorithms to carry out accurate 3D transfer of some of the principal cooling transitions via the application of a state-of-the art escape probability method developed by Poelman and Spaans (2005).

# References

[1] Ercolano B., Barlow M. J. , Storey P. J., Liu X.-W., 2003, MNRAS 340, 1136

[2] Ercolano B., Barlow M. J. & Storey P. J., 2005, MNRAS, 362, 103

[3] Gonçalves D., Ercolano B., Carnero A., Mampaso A. & Corradi R., 2005a, MNRAS, 365, 1039

[4] Poelman D. R. & Spaans M., 2005, A&A 440, 559

[5] Sahai R., Dayal A. et al., 1999, AJ 118, 468

[6] Speck A. K., Meixner M., Jacoby G. H., Knezek P. M., 2002, PASP 115, 170